# SECURE ROUTING PATH USING TRUST VALUES FOR WIRELESS SENSOR NETWORKS

Dr.S.Rajaram[1], A. Babu Karuppiah[2], K. Vinoth Kumar[3]

Department of ECE, Velammal College of Engineering and Technology

*ABSTRACT*

*Traditional cryptography-based security mechanisms such as authentication and authorization are not effective against insider attacks like wormhole, sinkhole, selective forwarding attacks, etc., Trust based approaches have been widely used to counter insider attacks in wireless sensor networks. It provides a quantitative way to evaluate the trustworthiness of sensor nodes. An untrustworthy node can wreak considerable damage and adversely affect the quality and reliability of data. Therefore, analyzing the trust level of a node is important. In this paper we focused about indirect trust mechanism, in which each node monitors the forwarding behavior of its neighbors in order to detect any node that behaves selfishly and does not forward the packets it receives. For this, we used a link state routing protocol based indirect trusts which forms the shortest route and finds the best trustworthy route among them by comparing the values of all the calculated route trusts as for each route present in the network. And finally, we compare our work with similar routing protocols and show its advantages over them.*

## 1. INTRODUCTION

A Wireless sensor network (WSN) is a collection of millimeter-scale, self-contained, micro-electro-mechanical devices. These tiny devices have sensors, computational processing ability (i.e. CPU power), wireless receiver and transmitter technology and a power supply. WSNs are limited in power, computational capacity and memory, and may not have global IDs. WSNs have a wide range of applications, ranging from monitoring environments, military zones, sensitive installations and remote data collection and analysis.

Trust establishment among nodes is a must to evaluate the trustworthiness of other nodes and is one of the most critical issues in WSNs. Trust is dependent on time; it can increase or decrease with time based on the available evidence through direct interactions with the node or recommendations from other trusted nodes. Trust-modeling is mathematical representation of node‟s opinion of another node in a network. We need mathematical tools to represent trust and reputation, update these continuously. Maintaining a record of the transactions with other nodes, directly as well as indirectly, from this record a „trust‟ value will be established [1].

Trust means everybody is trusted somehow and does not require any authentication (less overhead). It tells the degree of reliability. Every node finds the trust of all other nodes, based on previous experience and recommendations in fulfilling its promises.

## 2. SECURITY ISSUES IN WSNS

There are some successful trust schemes which can perform well such as Trust Management for Resilient Geographic Routing (T-RGR), Reputation-based Framework for Sensor Networks (RFSN) and Group-based Trust Management Scheme (GTMS), but several important problems still need to be solved.

(1) The first problem is, WSNs are made up of resources limited nodes. Energy is critical for

WSNs. Nevertheless, few of trust schemes take into consideration the node"s residual energy which can affect the work of the trust schemes. In general, when a node has a higher trust value in the sponsor node"s trust table, it may be apt to be selected as a service provider in a trust scheme. Then this can make the selected node consume more energy than other nodes and exhaust the energy prematurely. We learn that taking into consideration the node"s residual energy will avoid this flaw and prolong the network lifetime and balance the network load.

(2) The second problem is, when one node sends reputation request messages to its neighbor to get the monitored node"s indirect trust value, this can cost a large amount of additional energy.

Consider this case, in a group of n nodes, taking the RFSN trust model for an example let"s compute the amount of data packets that the nodes need to send out during getting indirect trust information. When node A wants to interact with node B, it needs to send out n-2 request packets and then receive n-2 replying packets, and the total number is 2(n-2). If node A wants to interact with n-1 nodes, the amount of packets is 2(n-1)(n-2).When all the nodes in the group update all trust value in their trust tables, so the maximum amount of the packets is 2n(n-1)(n-2). Now, we choose one node as a header, which is the trust information collection centre. The rest nodes in the group periodically report reputation messages to the header. The header collects and aggregates all the messages. The total number of the packets sent out in this process is n-1. When node A wants to communicate with node B, it just only need to send the reputation request message to the header and receives the reply message from the header. This will cost only 2 packets. If node A wants to communicate with n-2 nodes, there are 2(n- 2) packets all together. When all the nodes in the group update all the trust value of their neighbors, so the maximum amount of the packets is 2(n-2)(n-1)+n-1=(2n-3)(n-1). According to the energy consumption analysis proposed by H. O. Tan and I. Korpeoglu, once the size of packet and the distance are decided, the more number of packets are sent out, the more energy will be consumed. So the new method can cost less energy than the method described in RFSN, especially in the large-scale WSNs. At the same time, the header can aggregate the information, so it can reduce the memory consumption. And it can be easy to manage the trust of the mobile nodes as well. The new method may incur the header to exhaust energy quickly that is not beneficial for the network lifetime and the header will be easy to become the aim of the attacker. Altering the header properly and regularly may be a good method of solving this problem in this new method.

The third problem is, when the node integrates the indirect trust value from its neighbor nodes into the mixed trust value, it needs to pay attention on the reliability of the indirect trust value just received to avoid bad mouthing attack. In order to keep the WSNs secure, the node will receive recommendation information only from those nodes who have higher trust value in its trust tables, exceeding a predefined threshold (e.g. above 0.9). When it integrates all of the recommendation information, it allocates a high weight to the information sent from these nodes who have higher trust value. The higher trust value is, the higher weight it will be allocated. But there will be a flaw. Let us consider such kind of the node when it is selected to be a next routing node, it can service well, e.g. transmitting the received packet accurately. But when it is asked to provide the indirect trust information about the monitored node, it will send a false trust information to interfere the sponsor node to make accurate decisions. This kind of the node is called malicious spies[9]. These nodes can provide requested services and serve well, they are benign nodes in the sponsor node"s opinion. So the indirect trust information obtained from those malicious spies will be received as normal information. In RFSN and GTMS models, they all receive the introduction information from the nodes who has higher trust value. Obviously they are vulnerable to bad mouthing attack launched by malicious spies. To solve this defect, we first try to distinguish the confidence deposited in a node as a recommender and the trust deposited in the same node as a service provider [9]. But this method is not enough. We also consider to adopt filtering scheme that is used to filter interferential recommendation which has large difference with others in the all information and then punish or award depending on the filtering result. Isolating the malicious spies may consume more energy and resource. Maybe we can solve this problem by considering the benefit of the recommended nodes. We realize that the malicious spies will cost energy and bear the risk of being found as well as getting the benefit when they take participate in recommending. We can infer that when the cost is greater than income, it will not send out false indirect information; when the cost is less than income, it will send out false indirect information. This relationship between cost and income is a game relationship, so using game theory to solve the problem will be a good choice.

## 3. TRUST MODELING

A trust model can be defined as the representation of the trustworthiness of each node in the opinion of another node, thus each node associates a trust value with every other node.

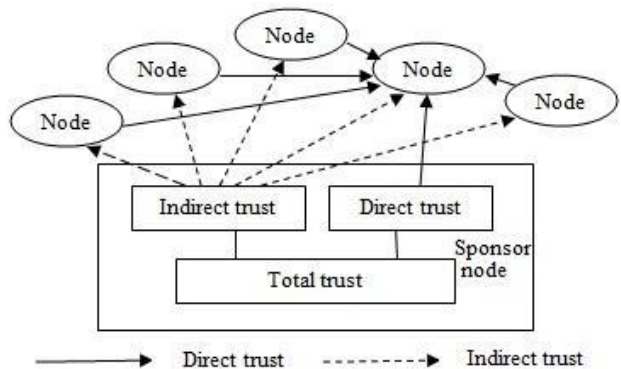

Figure 1. Combined Trust Mechanism

As illustrated in Figure, When the direct information can meet certain trust requirement, it uses the direct value, otherwise it will use the indirect trust value. Three different kinds information have their own advantages and disadvantages. Firstly, the direct trust value which is generated by the sponsor node itself is considered to be completely trustable and is the simple and energy saving. At the same time, it can prevent the bad mouthing attacks. On the other hand, the single direct trust value can make the information uncompleted and unsymmetrical. Sometimes the direct trust information can‟t accurately reflect the actual situation of the nodes. The indirect trust information is formed by the sponsor requiring and integrating. The indirect information alone is very rare in the WSNs. Omitting the indirect trust information can make the decision made depending on the trust value not be consistent with the actual state of the network[13]. Secondly, the indirect information can be used to initiate the new node‟s trust value[12]. We learn that the indirect information can help to solve the problem of the information uncompleted incurred by adopting the direct information alone. But requiring the indirect information can consume amount of additional of energy. This is critical for the sensor network lifetime and may bring the malicious nodes‟ bad mouthing attacks. We can infer that the direct information and the indirect information can supplement each other.

The updating of the trust value has two parts. The first one is the aging. We can give the different weights to the past trust value and the present trust value depending on the actual requirements and different destinations. If we give the present trust value a higher weight, it will always keep the node in the normal state. If we give the past trust value a higher weight, it will prevent the node‟s deception. Because the malicious nodes can work well in a short term to promote trust value. The second one is updating periodically. It can keep the routing lines stable and meet the characteristics of sensor nodes[14].The sensor nodes need to sleep in a timely manner to save energy.

## 4. RELATED WORK

Trust Management in WSNs is a relatively new concept which is still in nascent stage. Although some research papers are available for computing the trust value of the nodes, a very limited work towards their inclusion in the data routing path has been reported so far. In ATSR [1], the Total Trust (TT) information of a node is calculated by combining DT and IT information of the node. Finally the routing decisions are taken based on geographical information as well as TT information. The disadvantage is that the nodes needed to calculate DTs and ITs for all other nodes require high processing and storage capabilities. In DTLSRP [2], Total Direct Trust value (TDT) is calculated by geometric mean of all trust metrics. A threshold value (TTH) is fixed depending on application. The nodes having TDT higher than TTH are considered as benevolent nodes and rest are selfish nodes., Link State Routing protocol is used to find all available paths. The multiplicative Route Trusts are calculated for each path. The highest routing path trust value is chosen as best routing path. DTLSRP simulations perform better result with respect to other protocols such as ATSR, etc. This method allows us to find the shortest path without applying Dijkstra‟s algorithm. However it does not include

Indirect Trust (IT). TILSRP [3] is a new algorithm for calculation of both DT and IT of individual nodes based on LSR protocol. The selfish nodes are eliminated from the network and Route Trusts (RTs) are calculated thereby.

## 5. PROPOSED WORK

The indirect trust (IT) value is important mainly for newly initialized nodes or mobile nodes that have recently arrived in a new neighborhood. If a node is found in different neighborhoods each time it needs to send a new message, then direct interaction provides no trust info. In this case, if trust information is exchanged between nodes and especially if this exchange happens more frequently than the data message transmission, the source node has gathered (indirectly) trust information before it sends out a new data message. Thus, the exchange of indirect trust is useful mainly when node mobility has to be supported.

An important design option for calculating indirect trust information exchange is the frequency of the exchange. The exchange frequency directly affects the speed of trust information. To limit the exchanged information, only positive (or negative) information has been proposed to be shared [5]. However, when only positive information is shared, since nodes learn only from their own experience about a malicious node, colluding malicious nodes can extend each other"s survival time through false praise reports. If all N one-hop neighbors are asked triggering the generation of N reputation response messages, the network load would be significantly burdened (increasing collision probability) and the node resource (memory, processing and energy) consumption would also increase significantly.

From the analysis of direct trust mechanism, packet losses are severely occurred. So that we proposed one algorithm for obtaining secured routing path in WSNs. The algorithm is based upon both the direct and indirect trust exchange information. Both the information gives the secured and reliable path for routing. In this model, every node monitors all one-hop neighbors to check whether they forward the messages they receive towards the sink node. The ratio between the actually forwarded and all the transmitted messages provides an indication, whether the selected neighbor is sincerely cooperating. Values close to1 indicate a benign node, while values close to 0 indicate malicious nodes. This information is then used to calculate the routing function, based on which the next hop node is selected.

### 5.1 EVALUATION OF INDIVIDUAL TRUST METRICS IN SENSOR NODES

Suppose we take 23 wireless sensor nodes, to explain this algorithm. Although all the nodes are assumed to be homogeneous and uniform in hardware and routing capabilities initially, however their different parameters vary with time in a pre-defined manner in the example taken. Routing of sensed (or collected) data in this case can be initiated via event or time triggering. The method of routing, in the present case is developed independently of the rules associated with such triggering. Every packet of information to be routed definitely comprises of a source and a sink. They are denoted here by red and yellow colours. Streams of packets are sent from the Source (or N19) to the Destination (or N7) as shown in the figure below.

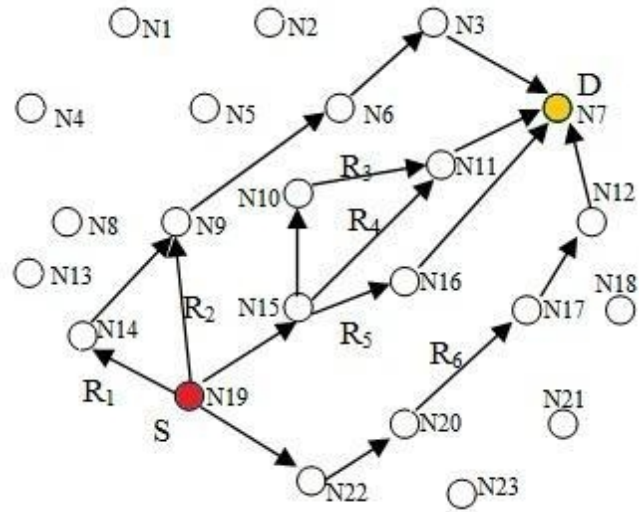

Figure 2. Routes from Source to Destination with node numbers as selected by LSRP

We are using the fundamental routing protocol named Link State Routing Protocol with the exclusion of Dijkstra"s algorithm. This protocol is particularly attractive in the case of Wireless Sensor Networks which have limited hardware and software features. The redundancy on applying Dijkstra"s algorithm here reduces the routing overhead. Its absence in calculation of the shortest path is compensated using our Select the Most Trusted Route (SMTR) algorithm which chooses the most reliable (or trusted) route. $R_1, R_2, R_3, R_4, R_5, R_6$ indicates the available routes from the source to destination.

$R_1$: S->N14->N9->N6->N3->D
$R_2$: S->N9->N6->N3->D
$R_3$: S->N15->N10->N11->D
$R_4$: S->N15->N11->D
$R_5$: S->N15->N16->D
$R_6$: S->N22->N20->N17->N12->D

**Algorithm:** For calculating the trust of the node.

**Initial condition:** Each node wants to communicate with other nodes in the network.

**Input:** Get the source and destination of the network.

**Output:** Trust value calculation and communication.

**Begin:** Apply Link State Routing Protocol.

- ❖ Get the Node id"s N=1

International Journal on Cryptography and Information Security (IJCIS), Vol. 4, No. 2, June 2014

- ❖ Mark the sensor nodes in each of the routes
- ❖ Calculate the direct trust by

$$DT_y(X) \text{ or } T_j(X) = \sum_{i=1}^{n} T_y(i)$$

- ❖ Calculate the indirect trust by

$$IT_i(X) = \frac{\left[\sqrt{(\frac{1}{N}\sum_{x=1}^{N}(T_j(X) * W_j)) * ((\frac{1}{M}\sum_{y=1}^{M}(T_j(X) * W_j))} \right] + [\frac{1}{P}\sum_{z=1}^{P}(T_j(X) * W_j)]}{2}$$

- ❖ Calculate the total trust by

$$TT = DT_y(X) + IT_i(X)$$

- ❖ Get the path trust by

$$PT = \prod_{w=1}^{n}(Rw)$$

- ❖ Select route N
- ❖ If PTN>PTi for all i  N
      Path through Route N
- ❖ Else
      N=N+1 and go back to previous step

**Notations:**

$DT_y(X)$: Direct trust value of node Y in node X

$IT_i(X)$: Indirect Trust value of Node X with respect to node i

$W_j$: Weightage given to the direct trust with respect to particular neighboring nodes used in calculation of indirect trust

TT: Total Trust

PT: Path Trust

## 5. SIMULATION RESULTS

To verify the performance of new algorithm, simulation is done through MATLAB.

International Journal on Cryptography and Information Security (IJCIS), Vol. 4, No. 2, June 2014

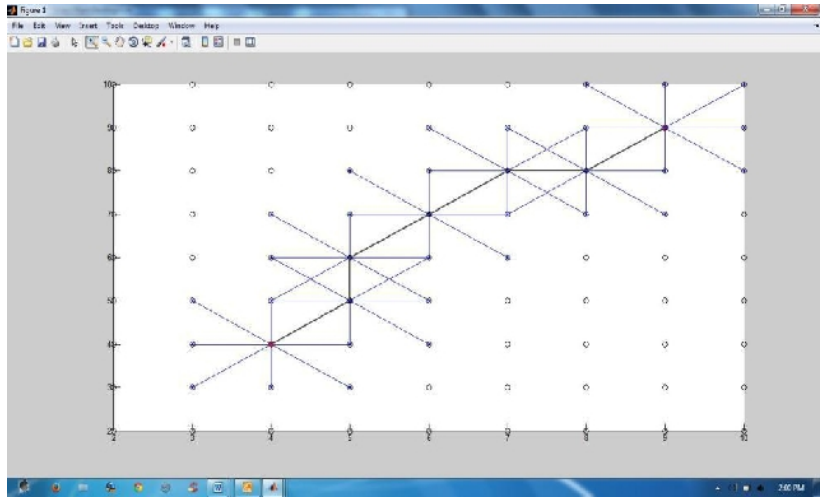

Figure 3. Trusted path for Routing from analysis

There are 50 nodes randomly uniformly distributed in the simulation environment, and the area is a circle with radius of 100 meters. The initial energy of each node is 5J and about every 1 second, each node generates a data packet with packet size 500bit. Here equivalent trust values of 0.5 were given in prior to some 30 fully powered and working homogeneous nodes and they sent messages in proportional to those value. It is clear from the fact that in absence of any receiving or sending nodes the indirect trust value of a node is extremely small. Then the indirect trust value is 0. The matlab simulation and performance analysis of this model is given above figure 3.

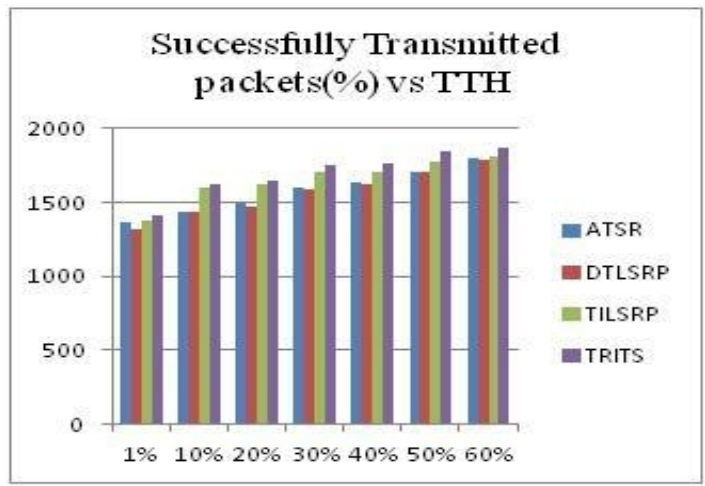

Fig 4. Performance analysis of TRITS compared with ATSR, DTLSRP and TILSRP

Our proposed method TRITS has better performance as compared to the existing systems ATSR, DTLSRP and TILSRP as shown in above figure 4.

34

## 6. CONCLUSIONS

Taking routing as main objective, proposed routing mechanism dedicated for wireless sensor networks. In our proposed method, the new algorithm SRPT (Secure Routing Path using Trust values) has better performance as compared to existing systems. Here, in this approach, during transmission of packets, if any node in the routing path get fails to transmit the packets. That time it can automatically chooses the another routing path to transmit the packets to the required destination.

**Authors**

**Dr. S. Rajaram,** working as Associate Professor of ECE Department of Thiagarajar College of Engineering, Madurai, obtained his B.E., degree Thiagarajar College of Engineering from Madurai Kamaraj Univeristy, Madurai and M.E. degree from A.C.C.E.T Karaikudi. He was awarded PhD by Madurai Kamaraj University in the field of VLSI Design. He was awarded PDF from Georgia Institute of Technology, USA in 3D VLSI. He has 16 years of experience of teaching and research. His area of research includes VLSI Design, Network Security, Wireless Networks. He has published many research papers in International journals, national and international conferences. 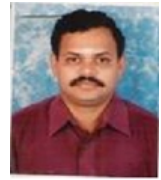

**A. Babu Karuppiah**, Assistant Professor of ECE Department of Velammal College of Engineering & Technology, Madurai, obtained his B.E., degree from Adhiparasakthi Engineering College, Madras University and M.E. degree from Mepco Schlenk Engg. College, Sivakasi that is affiliated to Anna University, Chennai. He has 11 years of Teaching and Research experience. Pursuing PhD in Anna University, Tirunelveli in Networking. He published many research papers in International journals, National and International conferences. His area of research includes Wireless Sensor Networks. 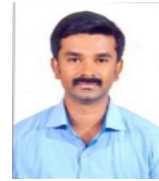

**K. Vinoth Kumar** pursuing final year ME Communication Systems at Velammal College of Engineering and Technology. He received the B.Tech degree from the Kalasalingam University in 2012. He is currently doing her project work in Wireless Sensor Networks and has published papers in International journals, national and international Conferences. 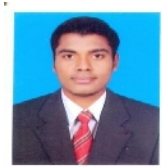